# Easily tunable long photonic hook generated from Janus liquids-filled hollow microcylinder

Zeng Peng,[1,2] Guoqiang Gu,[1,3] Liyang Shao,[1,4] Xingliang Shen[1]

[1]*Department of Electrical and Electronic Engineering, Southern University of Science and Technology, Shenzhen 518055, China*
[2]*Harbin Institute of Technology, Harbin 150001, P. R. China*
[3]*guguoqiang2009@gmail.com*
[4]*shaoly@sustech.edu.cn*

**The photonic hook, a beam that can propagate along a curved path, has attracted wide attention since its inception and experimental confirmation. In this paper, we propose a new type of structure, which was made by a hollow microcylinder and a Janus-shaped liquid column of two insoluble filling liquids, for producing photonic hook of easily tunable properties and long length. The $E^2$ field intensity distribution characteristics and formation mechanism of the photonic hook are studied by analyzing the energy flow using the finite element method. The profile and properties of the photonic hook can be effectively tuned by rotating the hollow microcylinder or changing the light incident angle. A long photonic hook with a decay length of ~18λ and a photonic hook with a large focal distance ~8λ are obtained by this model.**
© 2020 Optical Society of America

http://dx.doi.org/10.1364/OL.99.099999

As the basic point of geometric optics, light travels along a straight line in a homogeneous medium, which has been well known since ancient times. This view is also supported by the great Maxwell electrodynamics to ensure the conservation of momentum [1]. Until 1979, in the context of the maturation of quantum mechanics, this deep-rooted view was broken by Berry and Balazs [2]. For the first time, they solved the wave packets of the Airy beam in Schrodinger's equation, which has the properties of no diffraction in free space. In 2007, the curved Airy beam was first confirmed in experiments by Siviloglou et al. [3]. The Airy beam of which can propagate along a curved trajectory shows great potential in the fields of light capture, manipulation of nanoobjects, electron acceleration, and super-resolution imaging [4]. However, the Airy beam is often generated by bulky static phase plates or dynamic diffraction devices [3].

In 2016, I. V. Minin and O. V. Minin, et al. first proposed and validated that the symmetry-broken microparticles constituted by cuboids and prisms can produce another new kind of bending beam, the so-called photonic hook [5]. The formation of the photonic hook is much simpler than Airy beams [6], which is potentially increasing the possibility of extending the excellent application of curved beams to on-chip laboratories, such as optical-mechanical manipulation of nanoparticles [7-9]. Liu et al. first proposed a photonic hook in reflection mode achieved by the dielectric-coated concave hemicylindrical mirror, which could be the potential for integrated photonic circuits [10]. The idea of photonic hook has also been extended to acoustical hooks [11], and one application of acoustic hooks to trap particles was studied [12]. In previous work, twin photonic hooks were proposed for its great potential in manipulation with multiple particles [13, 14]. Likewise, designing a long optical hook is also essential for the efficient manipulation of nanoparticles, especially for moving nanoparticle in a long distance.

In a sense, the photonic hook is a kind of photonic nanojet produced by the microparticles with asymmetric structures [6, 7], asymmetric material compositions [15] or asymmetric illumination [16]. In contrast, the remarkable characteristic of the formation of traditional photonic nanojet is symmetry. First, a traditional photonic nanojet is formed at the shadow side of a symmetrical particle, which can be divided into two parts along the direction of the incident light, and these two parts are symmetric relative to the cutting face. Second, plane electromagnetic waves symmetrically irradiate the particle, which means the illumination region is symmetric relative to the entire structure [17]. In this paper, a kind of symmetric geometry structure but with asymmetric dielectric composition, which contains a hollow microcylinder and two insoluble liquids that can form a liquid-liquid interface, is designed to produce an easily tunable and long photonic hook. By rotating the hollow microcylinder, the angle between the liquid-liquid interface and incident light will be changed. Therefore, the properties of the photonic hook can be adjusted, including the curvature and full width at half maxima (FWHM) of the photonic hook. The simulation results based on the energy flow distribution and the corresponding Poynting vector diagram illustrate the formation mechanism of the photonic hook. Furthermore, a long photonic hook with the decay length of ~18λ is obtained utilizing a Janus liquids-filled hollow microcylinder with the rotation angle $\xi = 33°$.

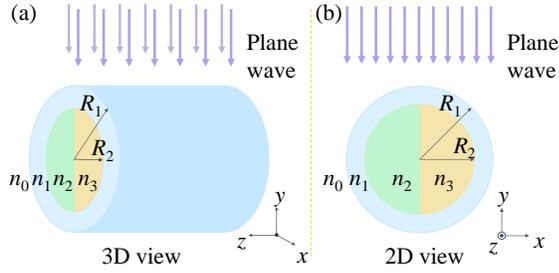

**Fig. 1** Schematics of plane-wave-illuminated Janus liquids-filled hollow microcylinder: (a) 3D stereogram and (b) 2D sectional view.

We mainly used the frequency-domain electromagnetic wave module of the finite element method based COMSOL Multiphysics commercial software package on studying the Janus liquids-filled hollow microcylinder. Figure 1(a) shows a three-dimensional (3D) schematic diagram of the hollow microcylinder in Cartesian coordinate system. The hollow microcylinder can be considered infinitely long in the $z$-axis direction. Therefore, a two-dimensional (2D) model was studied instead of a 3D model. Figure 1(b) shows the 2D cross-section diagram of the investigated model. The hollow microcylinder with refractive index $n_1 = 1.5$, outer diameter $R_1 = 4$ μm, and inner diameter $R_2 = 3$ μm, are placed in the air (with refractive index $n_0 = 1$). Two liquids filled in the hollow microcylinder have refractive indices of $n_2$ and $n_3$, respectively. A stable liquid-liquid interface will be formed between two immiscible liquids if interfacial tension is enough [18, 19]. The incident beam is a monochromatic plane wave with a wavelength $\lambda = 400$ nm traveling along the opposite direction of the $y$-axis. The perfectly matched layer absorbing boundary condition is set around the computational domain to avoid undesired back reflections.

Figure 2(a) shows the $E^2$ field intensity distribution diagram when only one liquid with a refractive index $n_2 = 1.41$ is injected into the hollow microcylinder. The light field with the characteristics of symmetric distribution relative to the optical axis is generated on the shadow side of the hollow microcylinder, that's the traditional photonic nanojet. Figures 2(b) and 2(c) show the $E^2$ field intensity distribution diagram when the incident light perpendicular to the liquid-liquid interface, the refractive indices of liquids combinations along the direction of incident light are $n_2 = 1.41$, $n_3 = 1.33$ and $n_2 = 1.33$, $n_3 = 1.41$, respectively. The $E^2$ field intensity distribution is still symmetrical to the incident light transmission axis. When the incident light goes through the liquid with a low refractive index and then goes through the liquid with a high refractive index, a longer photonic nanojet is formed. As shown in Fig. 2(d), the refractive indices of the two liquids are also $n_2 = 1.33$, $n_3 = 1.41$, which are the same with the values given in Figs. 2(b) and 2(c), but the incident light transmission axis is parallel to the liquid-liquid interface. In this case, the light field generated on the shadow side of the hollow microcylinder is no longer axisymmetric but curved with a significant angle, which means a photonic hook is formed.

According to the classical electromagnetic theory, Poynting vector distribution and energy flow distribution can be used to illustrate the formation of photonic hooks [20-22]. The incident light polarizing in the $x$-axis and propagating along the $y$-axis is used to conduct two-dimensional electromagnetic simulation, the distribution of the Poynting vector (red arrows), and energy flow (red lines) is obtained in the $x$-$y$ plane. As shown in Fig. 3(a), when only one kind of liquid is filled in the hollow microcylinder, the Poynting vector and the energy flow distribution distribute symmetrically along the transmission axis (dotted line) relative to the whole computational domain. As shown in Figs. 3(b) and 3(c), the hollow microcylinder is filled with two kinds of liquids, and the incident light is perpendicular to the liquid-liquid interface. The incident light passes through the liquids' combination of $n_3 = 1.41$, $n_2 = 1.33$ and $n_2 = 1.33$, $n_3 = 1.41$, respectively. In Figs. 3(a), 3(b), and 3(c), the symmetric Poynting vector and energy flow converge into a focused beam as a typical photonic nanojet. In Fig. 3(d), the incident light propagation direction is parallel with the liquid-liquid interface. The liquids' refractive indices are different on both sides of the transmission axis. It can be seen that the Poynting vector and energy flow distribution is inhomogeneous on two sides of the transmission axis. The energy concentrated more on the right side, where the energy flow lines are denser and smoother than the other side. This is due to the liquids of the left and right sides have asymmetric dielectric properties, which result in the broken symmetry of wavefront phase distribution and $E^2$ filed distribution. Eventually, the curved beam is formed by the interference between beams from the left side and the right side, that is the photonic hook.

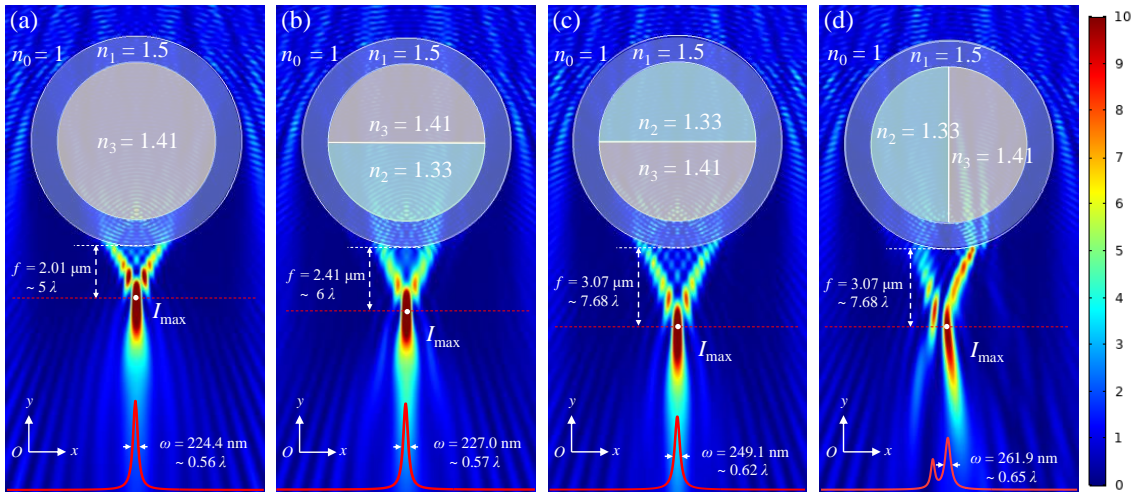

**Fig. 2** The $E^2$ field intensity distribution, focal distance $f$, and FWHM $\omega$ of the photonic nanojet formed by a hollow microcylinder with $R_1 = 4$ μm, $R_2 = 3$ μm and $n_1 = 1.5$, (a) filled by single liquid with $n_3 = 1.41$. (b) filled by liquid with $n_3 = 1.41$ and liquid with $n_2 = 1.33$, and incident light perpendicular to liquid-liquid interface. (c) filled by liquid with $n_2 = 1.33$ and liquid with $n_3 = 1.41$, and incident light perpendicular to liquid-liquid interface. (d) The $E^2$ field intensity distribution, focal distance $f$, and FWHM $\omega$ of the photonic hook formed by a hollow microcylinder with $R_1 = 4$ μm, $R_2 = 3$ μm and $n_1 = 1.5$, filled by liquid with $n_2 = 1.33$ and liquid with $n_3 = 1.41$, and incident light parallel to liquid-liquid interface.

We studied the varieties of waist and curvature of the photonic hook when the microcylinder rotates with angle $\xi$. The rotation angle $\xi$ is shown in Fig. 4(a), when the rotation direction is clockwise, the rotation angle $\xi$ is defined as positive. Conversely, when the rotation direction is counterclockwise, the rotation angle $\xi$ is defined as negative. The waist width is described by FWHM $\omega$ at the point of maximum light intensity. According to Refs. [10, 23], the curvature is described by the tilt angle $\alpha$ and bending angle $\beta$ of the photonic hook, which are determined by the start point, inflection point, and end point of the photonic hook, as shown in Fig. 4(b). The inflection point of the photonic hook is defined as the point with maximum light intensity, and the start and the end points are chosen at the positions where the field decays to $1/e$ of the maximum intensity along the direction of the photonic hook [15]. We get the clusters of points with light intensity decay to $1/e$ of the maximum light intensity, which can be divided into two parts. Along the forward curve of the photonic hook and choose the middle point as the start point on the arc that is near the microcylinder and almost perpendicular to the direction of the curve. The choosing of end-point is similar with the start point, but at another side of $I_{max}$ point, appearing at the end of the photonic hook. The first arm of the photonic hook is obtained by linking the start point and the inflection point, .and the second arm is obtained by linking the inflection point and the end point. The tilt angle $\alpha$ is determined as the angle between the first arm and vertical direction. The bending angle $\beta$ is the angle between the first arm and the second arm. In our study, the refractive index of hollow microcylinder is $n_1 = 1.5$, while the refractive index of the two-liquid filled in hollow microcylinder was chosen as $n_2 = 1.33$ and $n_3 = 1.41$ respectively. When the rotation angle changes from -90° to 90°, we obtained the variations of FWHM and bending angle. As shown in Fig. 4(c), we get photonic hook with maximum curvature at rotation angle $\xi = 33°$, whose bending angle $\beta = 150.3°$, tilt angle $\alpha = 23.6°$, and photonic hook with minimum FWHM $\omega = 0.425\lambda$ is obtained at rotation angle $\xi = 10°$.

To illustrate the length of photonic hook, a decay length $L$ is defined as the distance from the point of the maximum light intensity to $1/e$ of $I_{max}$. The photonic hook with a longest decay length $L$ of ~$18\lambda$ is obtained at rotation angle $\xi = 33°$, as shown in Fig. 5(a), which is the longest photonic hook comparing with the photonic hook already reported. This kind of long photonic hook also has large curvature of bending angle $\beta = 150.3°$ and tilt angle $\alpha = 23.6°$. Photonic hook with longest focal distance $f \sim 8\lambda$ is obtained in the case of $\xi = 40°$, as shown in Fig. 5(b). Figures 5(c) and 5(d) show that the FWHM $\omega$ of this kind of photonic hook is sub-wavelength level. From Fig. 5(c) and 5(d), we can also find that the $E^2$ field intensity distribution along the transverse direction at the point of maximum light intensity has a large gradient, which is potential for nanoparticle trapping and manipulation.

In this model, the refractive index of the hollow microcylinder is greater than the refractive index of the background medium (air) and the refractive indices of the two liquids filled inside. From a geometric optics perspective, the total internal reflection may occur when light waves propagate from optically dense media to thinner media. Therefore, the propagating beam from the solid microcylinder has the characteristics of rapid convergence and divergence. When liquids filled in hollow microcylinder, wider exit beam position distribution and a smoother exit ray will result in a longer attenuation length $L$ and a longer focal distance $f$ [24], which leads to the formation of a long photonic hook.

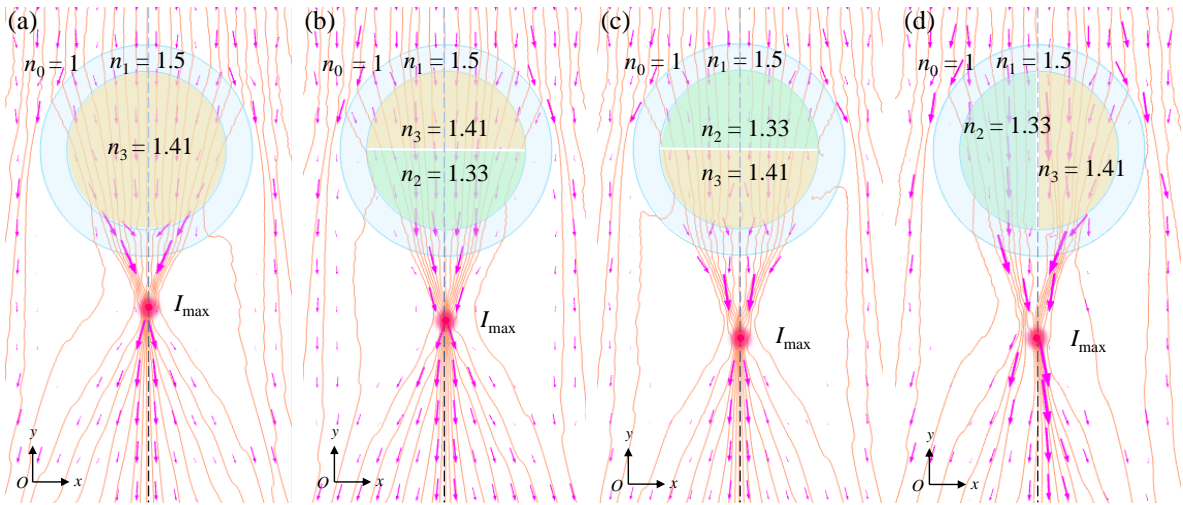

**Fig. 3.** Poynting vector and energy flow diagrams of the photonic nanojet formed by a hollow microcylinder with $R_1 = 4$ μm, $R_2 = 3$ μm and $n_1 = 1.5$, (a) filled by single liquid with $n_3 = 1.41$. (b) filled by liquid with $n_3 = 1.41$ and liquid with $n_2 = 1.33$, and incident light perpendicular to liquid-liquid interface. (c) filled by liquid with $n_2 = 1.33$ and liquid with $n_3 = 1.41$, and incident light perpendicular to liquid-liquid interface. (d) Poynting vector and energy flow diagrams of the photonic nanojet formed by a hollow microcylinder with $R_1 = 4$ μm, $R_2 = 3$μm and $n_1 = 1.5$, filled by liquid with $n_2 = 1.33$ and liquid with $n_3 = 1.41$, and incident light parallel to liquid-liquid interface.

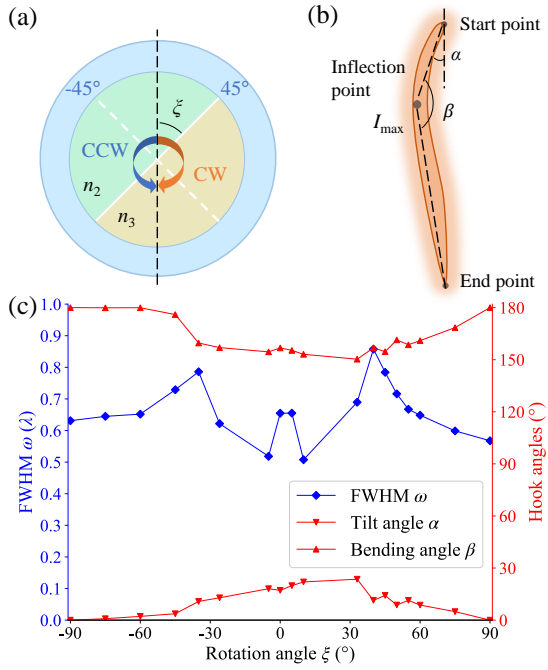

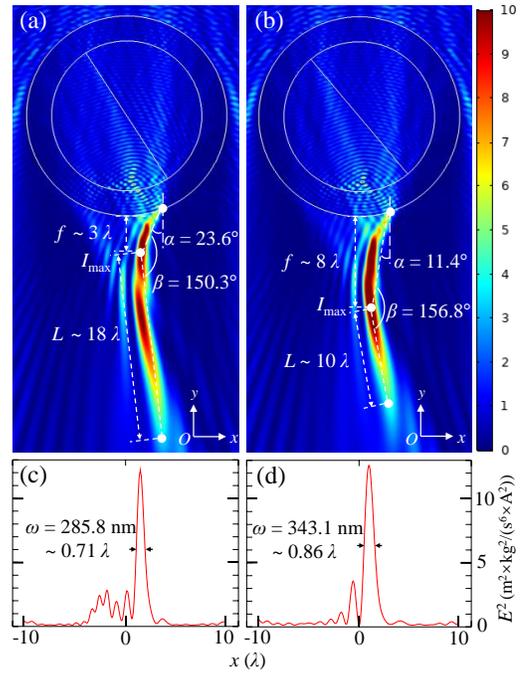

**Fig. 4** (a) Clockwise (CW) or counterclockwise (CCW) rotation of Janus liquids-filled hollow microcylinder with $n_2 = 1.33$ and $n_3 = 1.41$, $\xi$: rotation angle. (b) curvature definition of photonic hook, $\alpha$: tilt angle, $\beta$: bending angle. (c) FWHM $\omega$ and bending angle $\beta$ various with rotation angle from $\xi = -90°$ to $\xi = 90°$.

**Fig. 5** The $E^2$ field intensity distribution, decay length $L$, focal distance $f$, tilt angle $\alpha$ and bending angle $\beta$ for the cases of rotation angle (a) $\xi = 33°$ and (b) $\xi = 40°$. The field intensity profile along the transverse direction at the point of maximum light intensity and FWHM $\omega$ for the cases of rotation angle (c) $\xi = 33°$ and (d) $\xi = 40°$.

In conclusion, a new asymmetrical structure made by Janus liquids-filled hollow microcylinder, which is illuminated by a plane wave to produce photonic hook, is proposed via simulation using a numerical method. The simulation results show that the FWHM and bending angle of the photonic hook are easily tunable, which can be achieved only by rotating the hollow microcylinder or changing the light incident angle. Furthermore, long photonic hook with a decay length of ~$18\lambda$ is obtained at rotation angle $\xi = 33°$, and photonic hook with large focal distance ~$8\lambda$ is obtained at rotation angle $\xi = 40°$. This work provides a new way to generate an easily tunable and long photonic hook, which provides new possibilities for nanoparticle manipulation and microfluidic platforms.

**Funding.** This work was supported by the Guangdong Basic and Applied Basic Research Foundation (2019A1515011242), Shenzhen Postdoctoral Research Grant Program (K19237504), the startup fund from Southern University of Science and Technology and Shenzhen government (Y01236228 & Y01236128).

**Disclosures.** The authors declare that there are no conflicts of interest related to this article.